\documentclass{appolb}
\usepackage{epsfig,amssymb,xspace}
\newcommand{\nc}{\newcommand}
\newcommand{\order}[1]{ \mathcal{O} \left( #1 \right) }

\newcommand{\mathf}{\mathcal{F}}
\newcommand{\ave}[1]{\left\langle #1 \right\rangle}

\newcommand{\eqcomma}{\phantom{A},\phantom{A}}
\nc{\req}[1]{Eq.\,(\ref{#1})}  \nc{\eps}{\varepsilon}
\nc{\beq}{\begin{equation}}     \nc{\beql}[1]{\begin{equation}\label{#1}}
\nc{\eeq}{\end{equation}}       \nc{\rf}[1]{figure  \ref{#1}}
\nc{\beqa}{\begin{eqnarray}}   \nc{\eeqa}{\end{eqnarray}}
     \nc{\pathlaptop}{/home/rafelski/figure/}
     \nc{\pathletes}{/users/lpthe/jletes/bookraf/figures/}
     \nc{\pathnow}{}
 \def\lessim{\lower.5ex\hbox{$\; \buildrel < \over \sim \;$}}
\def\gtrsim{\lower.5ex\hbox{$\; \buildrel > \over \sim \;$}}

\usepackage{graphicx}

\begin{document}
\title{Relativistic fluid dynamics and its extensions as an effective field theory}
\author{David Montenegro$^a$ Radoslaw Ryblewski$^b$, Giorgio Torrieri$^a$
\address{$^a$IFGW, State University of Campinas, Brazil\\
$^b$ Institute of Nuclear Physics Polish Academy of Sciences, PL-31342 Krakow, Poland}}
\maketitle
 \begin{abstract}
We examine hydrodynamics from the perspective of an effective field theory.  The microscopic scale in this case is the thermalization scale, and the macroscopic scale is the gradient, with thermal fluctuations playing the role of $\hbar$.
We argue that this method can be applied both, to consistently include thermal fluctuations in the theory, and to extend hydrodynamics to systems whose microscopic structure is non-trivial.   For the latter, we discuss the case of spin polarization and gauge theories.
 \end{abstract}
\section{Introduction}
The problem of how to obtain, from first principles, a macroscopic description of a many-body system obeying a particular microscopic theory is a formidable one.   

Statistical mechanics techniques provide a unique and straightforward link via the thermal partition function assuming the system is in equilibrium.
Departure from equilibrium, however, requires extra assumptions about how the dynamics links degrees of freedom (DoF) at different scales.

The most common approach to deal with this issue in other contexts is effective field theory (EFT):  one constructs the most general lagrangian by identifying the symmetries of the microscopic theory, and expanding around the ratio of scales between the microscopic and the macroscopic DoF.
In the ``bottom up'' version of this approach, only macroscopic DoF are used to construct the theory.

Within field theory, such techniques have been applied to a variety of quantum and classical systems.
Statistical mechanics, however, gives new challenges:  typically, such systems are dissipative, probabilistic and non-linear, precluding the general applicability of unitary evolution and fluctuation-dissipation approaches. Typically, these theories have many scales, which do not always combine in a straightforward manner, and resolving these issues without reference to microscopic DoF is not always simple.

A case in point is relativistic hydrodynamics and transport theory.
The most common approach so far has been to obtain the former from a quasi-particle picture via the Boltzmann equation, or, alternatively, from a higher dimensional version of general relativity using holographic techniques.
Such approaches have yielded a considerable amount of activity and phenomenological success.   What is sometimes overlooked is that these are ``top-down'' approaches, where the microscopic theory is known explicitly.

In the case of the Boltzmann equation, it is itself a truncation of a tower of equations, called the BBGKY hierarchy in classical theory, which keeps track of all possible correlations of microscopic DoF.   The Boltzmann equation, via the assumption of molecular chaos, assumes such microscopic correlations become irrelevant, and only macroscopic correlations (caused by macroscopic dynamics) survive.   In the quantum version of the BBGKY hierarchy, higher order correlations invariably enter both thermal fluctuations (which are considered part of the macroscopic DoFs) and higher order perturbation theory (which is part of the microscopic effective lagrangian), producing micro-macro correlations which are never taken into account.
Holographic methods actually rely on a similar hierarchy via the large $N_c$ expansion, which typically suppresses correlations between multiple microscopic particles.
Physically, this truncation means that since there are ``very many'' DoF, correlations between them must be irrelevant.

Experimentally, however, we find systems close to the ideal hydrodynamic limit close to 50 particles, where such a reasoning is suspect.
Additionally, one may face a problem if the microscopic DoF have an ``internal structure'' or non-local symmetries.
For example, if macroscopic DoF carry spin ( something at the heart of the currently topical issue of ``chiral transport'') or obey non-local symmetries (magnetohydrodynamics), effective theory expansions are not uniquely defined.   Typically, there are three length scales at play
\begin{equation}
 l_{micro} \ll l_{mfp} \ll l_{macro}.
\end{equation}
Here $l_{mfp}$ is the dissipative mean free path, and the second inequality is the one mostly considered. On the other hand, $l_{micro}$ is the microscopic scale that knows about fluctuations and correlations, due to internal structure or gauge symmetries, and is usually neglected.

In this work, we suggest a way to construct a bottom-up approach that could resolve these ambiguities.   The idea is to minimize the assumptions regarding ultraviolet theory, beyond the fact, assumed to be true, that it reaches approximate local equilibrium.    We then apply this theory to various extensions of hydrodynamics, such as the addition of chemical potential, polarization, and gauge symmetry.      Section \ref{gene} will summarize Ref.~\cite{paper1}, while section \ref{ext} will summarize, subsequently Refs.~\cite{paper2,paper3,paper4} in section \ref{spin} and Ref.~\cite{paper5} in section \ref{gauge}.
\section{The general theory \label{gene}}
\begin{figure}[t]
\begin{center}
\includegraphics[height=0.17\textheight]{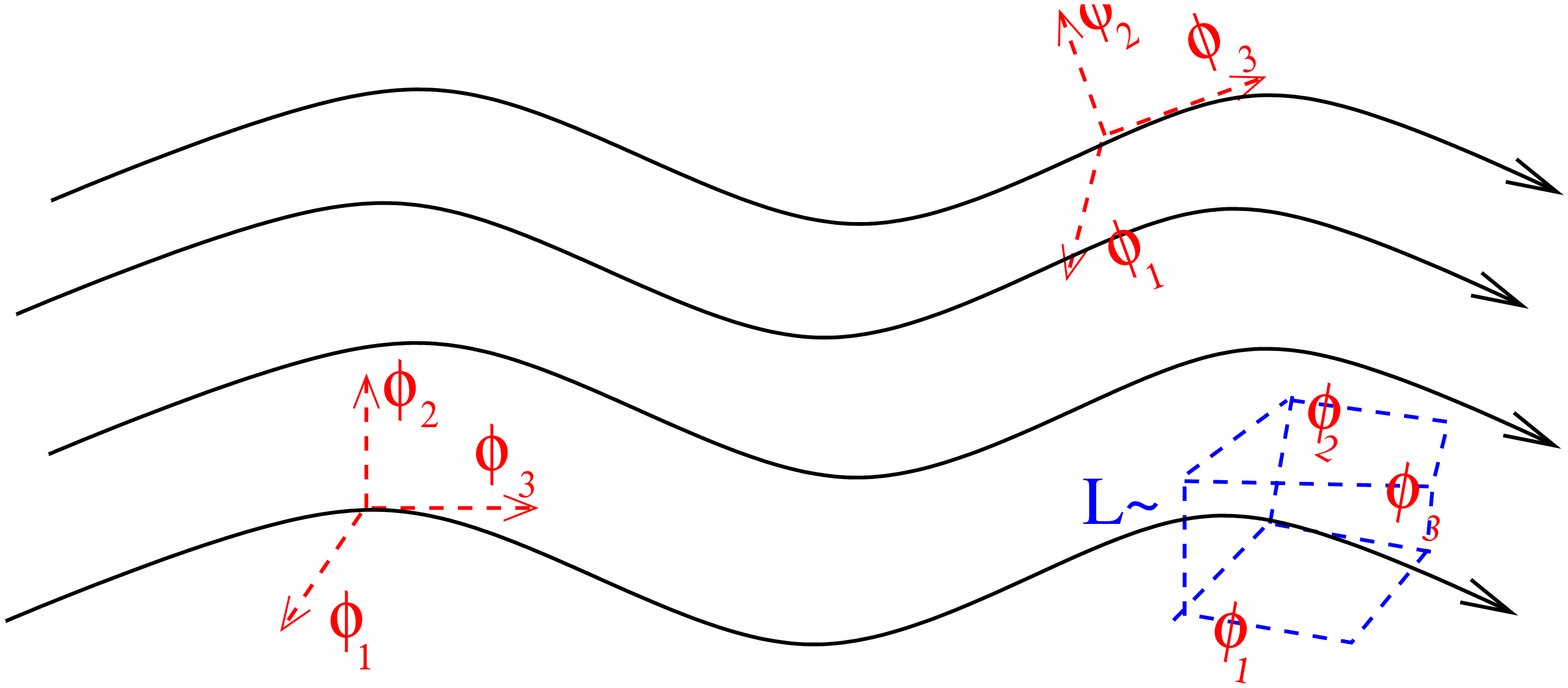}
\caption{\label{como}A representation of a fluid as a field, with the dynamics dependent on each cell volume
}
\end{center}
\end{figure}

We isolate two ``EFT scale expansion'' small parameters:
\begin{description}
\item[$\alpha$] is the usual Knudsen number giving the dissipative term attached to a gradient operator,
\item[$\beta$] is the relaxation term, and
  \item[$\gamma$] is the microscopic fluctuation term,
\end{description}
and the only assumptions we make are:
\begin{description}
\item[(a)] the existence of a locally isotropic vacuum carrying a well-defined amount of entropy which is conserved in the vanishing gradient stage,
\item[(b)] the stability of this vacuum against linearized perturbations, and\\
\item[(c)] a gradient expansion based on scale separation.
\end{description}
Let us examine in detail the effect of each assumption.   Assumption $(a)$ says that any dissipative DoF must be orthogonal to the velocity vector.
This means that in the absence of large gradients, there must be a local (diffeomorphism-invariant) $SO(3)$ coordinate symmetry and a global volume-preserving symmetry (which ensures entropy is a conserved quantity).   Thus, the zeroth order in gradient DoF are Lagrangian coordinates $\phi_{I=1,2,3}$ and dynamics can only be determined by an arbitrary function of
\begin{equation}
B = {\rm det}_{IJ} \, \partial_\mu \phi^I \partial^\mu \phi^J,
\end{equation}
that is the Jacobian of the volume-invariant transformations.  This is the most general ideal-limit term.   The rest will be the most general dissipative terms.
Thus, we can think of $\partial_\mu \phi_I$ as the Vierbein of a general coordinate transformation between the $SO(3)$-invariant comoving frame and a Lorentz-invariant laboratory frame.  This is illustrated in Fig. \ref{como}.
Dissipative terms must of course break these general coordinate symmetries, but the existence of a vacuum invariant under them puts constraints on this theory.  These definitions straightforwardly lead to the definition of the flow velocity $u^\mu$ as $u^\mu \partial_\mu \phi^I=0$, i.e.
\begin{equation}
 K_\mu = \frac{1}{6} \epsilon_{\mu \alpha \beta \gamma} \epsilon_{I J K} \partial^\alpha \phi^I \partial^\beta \phi^J \partial^\gamma \phi^K \eqcomma u_\mu = \frac{K_\mu}{\sqrt{B}}.
  \end{equation}
The most obvious assumption consistent with (a) is that additional terms must in general be orthogonal to the velocity of the fluid element (non-orthogonal components would renormalize the entropy).
The most general form of such additional DoF is
\begin{equation}
\label{genterms}
  \pi^{\alpha_1 \beta_1...\alpha_n \beta_n} = \partial^{\alpha_1} \phi^{I_1} \partial^{\beta_1} \phi^{J_1}   ... \partial^{\alpha_n} \phi^{I_n} \partial^{\beta_n} \phi^{J_n}  X_{I_1 J_1.. I_N J_N}
  \end{equation}
where $X_{I_1 .. I_N}$ are new DoF defined in the co-moving frame (a generalization of Israel-Stewart approach)  and  $\partial^{\alpha_i} \phi^{I_i}$ ensures flow orthogonality.
The fact that there is an infinite tower of such new DoF is ensured by assumptions (b) and (c), and Ostrogradsky's theorem:  Each new gradient arising from must be an asymptotic state to which the new degree of freedom evolves to.     Ostrogradsky's theorem together with (b) and (c) actually implies a stronger constraint
\begin{description}
\item[(b')] any linearized perturbations must be reconducible to ``entropy perturbations''
\end{description}
which ensures all $X$'s are purely dissipative and will have further applications in section \ref{spin}.

At first, it appears suprising that this infinite tower increases in tensor rank, $X_{IJ},X_{IJK},X_{IJKL},...$.   The necessity of this can be intuited from Ref.~\cite{transient}, where the ``Reynolds number'' is also introduced as the expansion parameter in tensor rank, in general distinct from the Knudsen number.
As can be seen in Ref.~\cite{transient}, the Boltzmann equation will have an infinite number of DoF corresponding to higher-order moments
\[\ \int d^4 p \,d^4 x\, p^{\alpha_1} p^{\alpha_2}...\,p^{\alpha_{2n}} f(p,x) . \]
These will have a more and more elaborate tensor rank, and their own equation of motion which will impact the evolution of any coarse-grained dynamics.
The fact  that each additional term in this tower comes in pairs (here labelled $\alpha_i,\beta_i,I,J$, thereafter just $\alpha_{1...2n},I_{1...2N}$ for brevity) is a consequence of the fact that the energy-momentum tensor is of rank 2.    The maintenance of the $\pi_{\alpha_1...\alpha_{2n}}$ tensor as a symmetric tensor in different $\alpha,\beta$ (any contraction to rank 2 must be symmetric) precludes mixing between even-rank and odd-rank components.  In the particular case of the Boltzmann equation being the microscopic theory \cite{transient} one can see how the $\ave{\dots}$ makes sure that when chemical potentials are not included odd-rank tensors do not contribute to even-rank tensor equations of motion.

The expansion in Ref.~\cite{transient} (and any resummation of it!) will of course remain within the domain of the Boltzmann equation, and will miss the molecular chaos deviation driven by microscopic propagators $D(x_1,....,x_{2n})$ or $\tilde{D} (p_1,..., p_{2n})$
\[\ \int d^4 p_1 d^4 x_1...\,d^4 p_n d^4 x_n\, p^{\alpha_1} ...\, p^{\alpha_{2n}} f(p_1,x_1)... f(p_n,x_n) D(p_1,p_2,...| p_{2n-1} p_{2n}) . \]
Eventually such terms will overwhelm Boltzmann expansion terms, especially if (as is obviously true for systems with $N_c=3$ and $\order{100}$ DoF) the scattering length is of the order of the microscopic DoF separation.
These, however, will also have the same covariance structure and, assuming a hydrostatic vacuum, symmetry structure.   Hence, a Lagrangian built out of terms such as that in Eq.~\ref{genterms} will incorporate these terms.

Additional symmetries will of course add more constraints to this general covariance.   For example, the phenomenologically popular Bjorken solution would entail $X_{...1...}=X_{....2...}=X_{...3...}$, so the full DoFs are $X_1,X_{11},X_{111},...$, while Greek letters reduce to $\mu=0,3$ correlated by boost-invariance $u_3 = x_3/x_0,u_{1,2}=0$, so $\phi_{1,2}=x_{1,2},\phi_3=...$.

The assumption (b) of the hydrostatic limit implies the existence of a free energy to be minimized (i.e. an entropy to be maximized).  
It also forces the ``kinetic'' term of the lagrangian to give a dissipative equation of motion with the source given along Maxwell-Cattaneo lines (further corrections are of course possible but they are suppressed in the linear case).   This obviously means that we will have to set up our equation in the doubled DoF dissipative lagrangian formalism \cite{grozdanov} along the lines of a generalized Ref.~\cite{paper1}.

The full lagrangian will then be given by
\begin{equation}
\label{genlag}
  L= F(B) + \left[ \sum_{i=1}^{\infty} \sum_{j=1}^3 L_{i}^{j} \right] + \sum_{i_1,j_1,i_2,j_2} L_{i_1}^{j_1} L_{i_2}^{j_2}.
\end{equation}
The first class of dissipative terms, $\sum\limits_{i=1}^{\infty} \sum\limits_{j=1}^4$ are ``leading order'', necessary to generate the stable vacuum.  From Ref.~\cite{paper1} one has
\begin{equation}
  L_{i}^1 = \frac{1}{2} \order{\beta^i} \left[ \pi^{\alpha_1...\alpha_{2n}}_- u^\gamma_+ \partial_\gamma \pi_{ + \alpha_1...\alpha_{2n}} - \pi^{\alpha_1...\alpha_{2n}}_+ u^\gamma_- \partial_\gamma \pi_{ - \alpha_1...\alpha_{2n}} \right],
\end{equation}
\begin{equation}
  L_{i}^2 =  \frac{1}{2} \Pi_{\pm}^2,
\end{equation}
\begin{equation}
  L_{i}^3 =  \alpha^n \left[ K^\nu \partial^\mu \pi_{ + \alpha_1...\alpha_{2n-2}} \right]\partial_\mu \phi_I \partial_\nu \phi_J.
\end{equation}
The last term is not covariant but does not explicitly appear in the equations of motion, acting as a source term.   Each $\alpha_1,\alpha_2$ pair can of course be decomposed into a scalar and a tensor part, giving rise to generalizations of shear and bulk viscosity.

The second class of terms, in $\sum\limits_{i_1,j_1,i_2,j_2}$ are the most general EFT expansion generating terms which are both dissipative and non-linear.  They go away in the linearized limit.  These terms must be scalar in both greek and latin letters, and, to second order, they ensure that the Maxwell-Cattaneo fluid is generalized to Israel-Stewart form.
\section{Extensions \label{ext}}
\subsection{Chemical potential}
The most obvious extension to hydrodynamics is to add a conserved scalar charge, represented in field theory as a Noerther current for a global $U(1)$ symmetry.

The dynamics of conserved charges will have a similar structure to that examined in the previous section.  The only difference is that such charges will have relaxation terms already at first order, rather than with ``two gradients'', since in general the difference between the frame where the energy-momentum current is at rest from the frame where the charge is at rest (respectively the Landau and Eckart frames) provide dissipation.   Thus, to leading order the additional degree of freedom representing charge is determined by the chemical shift symmetry \cite{shift}, the mathematical representation of the fact that at equilibrium the charge current and the energy current will move in the same direction.  

If the generator of  the internal symmetry representing the conservation law is $\psi$, the dynamics to leading order will depend on $y=u_\mu \partial^\mu \psi$.   Beyond leading order chemical shift symmetry is broken, but in a way that the hydrostatic vacuum mantains it.  
In analogy to $\pi^{\mu \nu}$, the most general degree of freedom is \cite{chempot}
\begin{equation}
y^{\alpha_1 ... \alpha_n} = \partial^\alpha_1 \phi_{I_1}...\partial^\alpha_n \phi_{I_n} Y^{I_1...I_n}.
\end{equation} 
The $Y^{...}$ are only different from $X^{...}$ in that in $X$ each new term in rank start at rank one rather than two, since $T_{\mu \nu}$ is a tensor current and not a vector current.  For $Y_{...}$ tensor ranks come intervals of one.

Therefore the most general lagrangian given in Eq.~\ref{genlag} is generalized to
\begin{equation}
  L= F(B,y) + \left[ \sum_{i=1}^{\infty} \sum_{j=1}^3 \left( L_{i}^{j} + \mathcal{J}_i^j \right) \right] + \sum_{i_1,j_1,i_2,j_2} \left( L_{i_1}^{j_1} + \mathcal{J}_{i_1}^{j_1} \right) \left( L_{i_2}^{j_2}   + \mathcal{J}_{i_2}^{j_2} \right)
\end{equation}
where $\mathcal{N}_i^{1,2}$ are straight-forward extensions of $L$ with $\pi$ terms substituted by $J$ terms, while 
\begin{eqnarray}
\mathcal{N}_n^{3}= \alpha^n \left[  \partial^\mu j_{ + \alpha_1...\alpha_{n-1}} \right]\partial_\mu \phi_I.
  \end{eqnarray}
Notice that only one Vierbein is necessary here.
\subsection{Spin \label{spin}}
A more complicated internal symmetry is spin, since it is not explicitly conserved but summed together with angular momentum.   In the ideal fluid limit angular momentum is provided by circulation, a non-local quantity defined over a loop.

One can however still treat spin as a ``set of chemical potentials'', transforming under $SO(3)$ symmetry (averaged over many particles incoherently phases become irrelevant, so $SU(2)$ is equivalent to $SO(3)$), each chemical potential representing a polarization direction component.

A simple generalization of the chemical shift symmetry in these terms follows
\begin{equation}
\left. \Psi_{\mu \nu}\right|_{comoving} = -\left. \Psi_{\nu \mu} \right|_{comoving} =    \exp \left[ -\sum_{i=1,2,3}
\alpha_i(\phi_I)  \hat{T_i}^{\mu \nu}\right] 
\end{equation}
and
\begin{equation}
 \alpha_i \rightarrow \alpha_i+\Delta \alpha_i\left(
\phi_I \right) \Rightarrow L(b,y_{\alpha \beta} =u_\mu \partial^\mu
\Psi_{\alpha \beta} ),
\end{equation}
assuming that one rank of the rank-3  spin tensor is always proportional to flow $u_\mu$.
Ensures that polarization current (one index of the rank-3 tensor) always proportional to $u_\mu$.
\begin{figure}[t]
\begin{center}
\includegraphics[height=0.17\textheight]{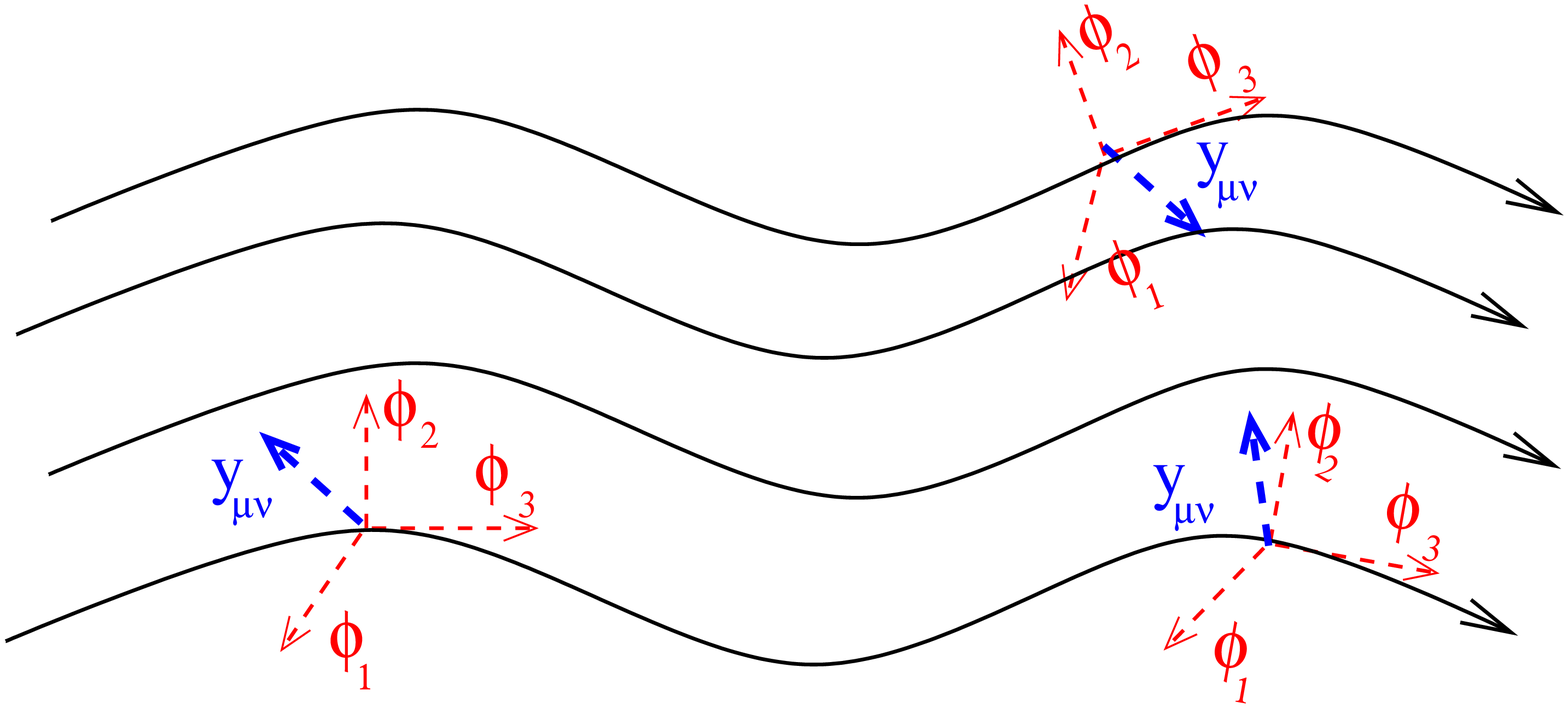}
\caption{\label{spinfield} The field theory of a fluid with spin as an auxiliary variable}
\end{center}
\end{figure}
If spin was conserved, every $\phi_I$ would aquire a $y_{\mu \nu}$ field, but this is not the case.  Instead, the generic equation of state for an ideal fluid with polarization is a function of $b$ and $y^2=y_{\alpha \beta} y^{\alpha \beta}$, as in Fig. \ref{spinfield}.  In the limit where polarization is small, one can always assume
\begin{equation}
L = F(b,y) = F\left( b \left( 1-cy^2 \right) \right),
\end{equation}
where a positive constant $c$ means a ferromagnetic material and a negative one an anti-ferromagnetic one.

\begin{figure}[t]
\begin{center}
\includegraphics[height=0.14\textheight]{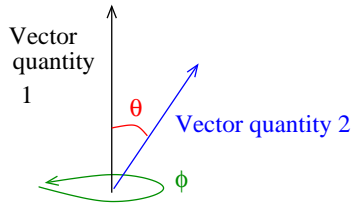}
\caption{\label{action} The reason why two dynamical pseudovector quantities not parallel to each-other will inevitably produce a Goldstone mode}
\end{center}
\end{figure}

What about the remaining two indices?   We are now ready to combine polarization with the ideal hydrodynamic limit, defined here as in the introduction:
\begin{itemize}
\item The dynamics within each cell is faster than macroscopic
dynamics, and it is expressible only in terms of local variables and
with no explicit
reference to four-velocity $u^\mu$ (gradients of flow are however
permissible and in fact required to describe local vorticity).
 \item Dynamics is dictated by local entropy maximization,
within each cell, subject to constraints of that cell alone.
Macroscopic quantities are assumed to be in local equilibrium inside
each macroscopic cell.
\item Only excitations around a hydrostatic medium are reducible locally to energy density perturbation.
\end{itemize}
The third point forces polarization and vorticity to always be parallel to avoid a Goldstone mode, as in Fig.~\ref{action}.
Equation of state is related to the Lagrangian in a way analogous to Ref.~\cite{paper3}. $L=\mathf$ where $\mathf$ is the Legendre transform of the Lagrangian w.r.t. polarization, hence
\begin{equation}
d \mathf = \frac{\partial \mathf}{\partial V} dV +
\frac{\partial \mathf}{\partial e}de + \frac{\partial \mathf}{\partial
\left[ \omega_{\mu \nu}\right]} d \left[\omega_{\mu \nu}\right]=0.
\end{equation}
The physical effect of polarization is to mix sound waves and vortices.  A linearization shows that such a lagrangian will always give unstable modes, which could be classified as Ostrogradsky instabilities.

An analogy with Israel-Stewart hydrodynamics, with spin {\em relaxing} towards vorticity according to 
\begin{equation}
\tau_\Omega  u_{\alpha} \partial^\alpha \Omega_{\mu \nu} + \Omega_{\mu \nu} =
  \chi(b,\omega^2) \omega^{\mu\nu} = y^{\mu\nu} ,
\end{equation}
derivable from a Lagrangian formulation in analogy to Israel-Stewart as
\begin{equation}
  \label{genlagr}
L = F(b(1-c\, y_{\mu\nu}y^{\mu\nu})) + \mathcal{L}_{IS - vortex},
  \end{equation}
\[\ \mathcal{L}_{IS - vortex} = \frac{1}{2} \tau_Y ( \, Y^{\mu\nu}_{-} \, u^{\alpha}_{+} \partial_\alpha Y_{\mu\nu +} - Y^{\mu\nu}_{+} u^{\alpha}_{-} \partial_\alpha Y_{\mu\nu -} ) + F^\prime(\alpha^3) \left\{ c \chi^2(b,w^2) \omega^{\mu\nu}\omega_{\mu\nu} \right\}.
\]
As shown in Ref.~\cite{paper4} such a prescription can restore causality if the relaxation time is long enough.   The price to be paid is a minimal amount of dissipation, which can be thought of as a ''minimal viscosity'' affecting all relativistic systems with a non-zero polarizability.

This discussion however leaves open the problem of Gauge invariance:  in a system where the microscopic theory is gauge invariant, the transverse polarization of the microscopic DoF vs the angular momentum is a gauge dependent quantity.   Hence, the free energy must be degenerate between vorticity, polarization and color space.   The consequences of this will be explored in the next section.
\subsection{Gauge symmetries\label{gauge}}
To understand gauge symmetries, let us start from Ref.~\cite{shift} and generalize this reasoning to a non-Abelian symmetry.   Both the phase and the current will have directions, and for the system not to be invariant under any rotations in internal space it must be that
\begin{equation}
  \label{gauggen}
y= J^\mu \partial_\mu \alpha \rightarrow \left[ J^\mu \right]_a  \partial_\mu \left[ \alpha \right]_b = y_{ab} .
\end{equation}
Note that normal chemcial potentials for $SU(3)$ invariant matter (isospin and charge) do not form a matrix.   This is because, while the Hamiltonian is invariant under that symmetry, the states live in different super-selection sectors and cannot mix.

\begin{figure}[t]
\begin{center}
\includegraphics[height=0.14\textheight]{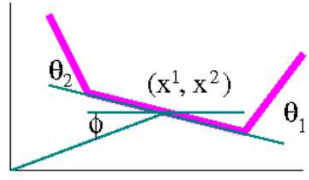}
\caption{\label{swimmer}A representation of the non-hydrodynamic DoF as ``swimming'' in the space of gauges}
\end{center}
\end{figure}

The formalism of Eq. \ref{gauggen} is however appropriate for locally equilibrated gauge fields, fluids where there is some ``red'',''green'' and ``blue'' charge density.   We should be able to re-parametrize this density {\em locally} without changing the dynamics of the fluid.    And it is easy to see that this must give rise to a free energy with a continuous landscape of non-local minima, similar to that of a protein \cite{folding}, where thermal fluctuations can radically change the macroscopic configuration of the fluid.
Mathematically, all we need to do to verify this is to impose invariance under the gauge symmetry.  Throughout
\begin{equation}
  \label{gaugedef}
F(y,...) = F(U^{-1}(x) y U(x)) \eqcomma U_{ab}(x) \in SU(N) = \exp \left[ \sum_i \alpha_i (x) \hat{T}_i  \right].
\end{equation}
Comparing Eq.~\ref{gauggen} and Eq.~\ref{gaugedef} one gets
\begin{equation}
  y_{ab} \rightarrow U^{-1}_{ac}(x) y_{cd} U_{bd}(x) = U^{-1}(x)_{ac} J^\mu_f U_{cf} U^{-1}_{fg}  \partial_\mu \alpha_{g} U_{bg} =
\end{equation}
\[\ =
  U^{-1}(x)_{ac} J^\mu_f U_{cf} \partial_\mu \left(  U^{-1}_{fg}  \alpha_d  U_{bd}(x) \right) - J^\mu_a \left( U \partial_\mu U \right)_{fb} \alpha_f.
\]
  The second term is impossible to satisfy without introducing additional microscopic dynamic variables
\begin{equation}
     F\left(b,J^\mu \partial_\mu \alpha \right) \rightarrow F\left( b, J^\mu \left( \partial_\mu - U(x) \partial_\mu U(x) \right) \alpha \right).
\end{equation}
The local equilibrium state, with both the gauge symmetry and the chemical shift symmetry, is
\begin{equation}
  \label{surprise}
 J^\mu_a = \frac{\partial F}{\partial y_a} u^\mu \eqcomma L= F(b,y_{ab} \left(1-u_\mu \partial^\mu \alpha^{i} ) \right)  \simeq F\left( b,Tr \left[ y_{ab} \left(1- (\hat{T}_{bc})_i u_\mu \partial^\mu \alpha^{i}  \right) \right]^2,... \right),
\end{equation}
where the last term, representing ``rotation'' in color space in the flow direction, is inherently non-hydrodynamical because it represents microstate redundancies.   We note that in electromagnetism, where no such rotations are possible, the problem does not present itself
\begin{equation}
  \hat{T}_i \rightarrow 1 \eqcomma y_{ab} \rightarrow \mu_Q \eqcomma u_\mu \partial^\mu \alpha_i \rightarrow A_\tau,
\end{equation}
but in non-Abelian theories, where twisting in gauge directions is allowed, no such redefinition is possible.
The ``non-hydrodynamic DoF'' can be physically thought of as ``swimming'' in the space of gauges.  It is well-known from non-relativistic hydrodynamics that force-free swimming is possible for bodies with a complicated enough topology (``purcell swimmers'', the swimmer analogy to falling cats).  The non-hydrodynamic modes can be thought of analogously, see Fig.\ref{swimmer}. 
When vorticity arises, the dynamics of the previous subsection becomes relevant, since
\begin{equation}
  \oint J_{i}^\mu dx_\mu   \equiv \int_\Sigma d \Sigma_{\mu \nu} \omega^{\mu \nu}  \ne 0 \rightarrow  \omega_{i}^{\mu \nu}= \epsilon^{\mu \nu \alpha \beta} \partial_\alpha J_{\beta ab} \ne 0
  \end{equation}
is not invariant under a gauge transformation, but has the transformation properties of a Wilson loop, the equivalent transformation in color space.
Thus, coupling the two will give a gauge-invariant term in the free energy, $Tr_{i} \left[ \omega_{\mu \nu} G^{\mu \nu} \right]$.
This term, in fact, is the vorticity-spin coupling examined in the previous subsection, since the infinitesimally short Wilson loop gives the electromagnetic field tensor which, in local equilibrium, gives the local polarization of the gauge bosons.
The issue with causality we examined in the previous section will also hold, but the ``relaxation minimum'' will have an additional degeneracy, giving non-hydrodynamic modes.
\section{Further developments}
So far, we have not touched the $\gamma$ constant, quantifying microscopic fluctuations.   This is because its role is similar to $\hbar$, quantifying deviations from classical physics.  It does not appear in the Lagrangian but will be noticeable in observables when random (quantum or thermal) fluctuations become non-negligible.
The effective energy-momentum tensor will then have the form
\begin{equation}
T^{\mu \nu}_{eff}=\frac{\delta \ln Z}{\delta g_{\mu \nu}} +  \hat{T}^{\mu \nu} \eqcomma J^\mu = \frac{\delta \ln Z}{\delta y_\mu } + \hat{J}^\mu
\end{equation}
where
\begin{equation}
\ln Z= \int \mathcal{D} \left[B,X\right] e^{T_0^4 \int \mathcal{L}(B,X) d^4 x}
\end{equation}
and $\hat{T},\hat{J}$ are stochastic source terms whose correlation is given by
\begin{equation}
\label{stoch}
\ave{\hat{T}^{\alpha_1 \beta_1} ...\hat{T}^{\alpha_n \beta_n}  }= \order{\gamma^n} \frac{\delta^n \ln Z}{\delta g_{\alpha_1 \beta_1} ... g_{\alpha_n \beta_n}} \eqcomma \ave{\hat{J}^{\alpha_1 } ...\hat{J}^{\alpha_n}  } = 
 \order{\gamma^n} \frac{\delta^n \ln Z}{\delta y_{\alpha_1} ... y_{\alpha_n }}
\end{equation}
with charge random sources having a similar structure.
The construction of a fluctuation-dissipation regime out of this system is a formidable but well-defined problem.     One would solve, deterministically, the equations of motion arising from the $\ln Z$ and at the same time use Eq.~(\ref{stoch}) to construct a source of stochastic fluctuations.   The combined effect of the classical evolution with the stochastic fluctuations would give the full evolution of the system.

The fact that vortices have no energy gap but do no propagation speed makes it obvious that this problem is highly non-perturbative, as is confirmed by explicity calculations in a deformed theory \cite{torrieri1}.  Polarization might fix this problem by introducting a soft dissipative "mass gap" that regulates the low amplitude ultraviolet modes.

Alternatively \cite{torrieri1,moore}, the Kolmogorov cascade, shifting low-frequency high amplitude modes into high-frequency low amplitude ones, {\em has} to end in a regime where frequency and amplitude of energy density coincide in natural units because of quantum considerations.   This is also a cutoff, but a highly non-perturbative one.

To study its effects, a lattice solution \cite{torrieri2} might be feasible.   Attempts to solve this problem perturbatively in a situation where vortices are neglected have also appeared in the literature \cite{fluct1,fluct2,fluct3} but, due to the empirical importance of vortices in turbulent systems, we think such calculations should be compared with ``realistic`` fluids only with extreme care.    The existence of the fundamental thermodynamic law  postulated in the beginning means an equilibrium state will arise from any initial condition, but the nature of it (the presence of macroscopic turbulence as well as microscopic thermalization, as suggested in \cite{torrieri2}?) and especially the timescale of equilibration, could be different from classical expectations.

In conclusion, we have given a broad overview of constructing dissipative hydrodynamics as a field theory, and tried to apply this formalism to extend the theory into domains such as spin and gauge theory.   This work is still very much under development, but at least some results of this approach appear well-defined.

 GT acknowledges support from FAPESP proc. 2017/06508-7, partecipation in FAPESP tematico 2017/05685-2 and CNPQ bolsa de
 produtividade 301996/2014-8.  This work is a part
 of the project INCT-FNA Proc. No. 464898/2014-5. RR  was supported in part by the Polish National Science Center Grants No. 2016/23/B/ST2/00717 and No. 2018/30/E/ST2/00432.

\end{document}